\pdfoutput=1

\documentclass[11pt]{article}

\usepackage{acl}

\usepackage{times}
\usepackage{latexsym}
\usepackage{natbib}

\usepackage[T1]{fontenc}

\usepackage[utf8]{inputenc}

\usepackage{microtype}
\usepackage{graphicx}
\usepackage{chngpage}
\usepackage{multirow}
\usepackage{float}
\usepackage{nccmath}

\newcommand{\blue}[1]{\textcolor{blue}{#1}}
\newcommand{\red}[1]{\textcolor{red}{#1}}

\title{Masked Audio Text Encoders are Effective Multi-Modal Rescorers}

\author{
Jinglun Cai, Monica Sunkara, Xilai Li, Anshu Bhatia, Xiao Pan, Sravan Bodapati \\ 
AWS AI Labs\\
 \small\texttt{\{cjinglun, sunkaral, xilaili, anshubha, panxx, sravanb\}}@amazon.com
 }

\begin{document}
\maketitle

\begin{abstract}
Masked Language Models (MLMs) have proven to be effective for second-pass rescoring in Automatic Speech Recognition (ASR) systems. In this work, we propose \textbf{M}asked \textbf{A}udio \textbf{T}ext \textbf{E}ncoder (\textbf{\mymethod}), a multi-modal masked language model rescorer which incorporates acoustic representations into the input space of MLM. We adopt contrastive learning for effectively aligning the modalities by learning shared representations. We show that using a multi-modal rescorer is beneficial for domain generalization of the ASR system when target domain data is unavailable. \mymethod reduces word error rate (WER) by 4\%-16\% on in-domain, and 3\%-7\% on out-of-domain datasets, over the text-only baseline. Additionally, with very limited amount of training data (0.8 hours) \mymethod achieves a WER reduction of 8\%-23\% over the first-pass baseline.

\end{abstract}

\section{Introduction}

Performance of Automatic Speech Recognition (ASR) systems has been traditionally improved during inference time via either editing/refinement \cite{fastcorrect_NEURIPS2021,chi-etal-2021-align,cai2023kgeco} or second-pass rescoring \cite{delibration_neurips17,sainath19_interspeech,hu2020deliberation} using language models . 
In recent studies, Transformer-based pre-trained Large Language Models (LLMs) have shown promising results when used as second-pass rescorers.  Previous works \cite{rescoreBERT, salazar-etal-2020-masked, llm_22interspeech} have shown that deep bidirectional Transformers \cite{devlin-etal-2019-bert} perform better than their unidirectional counterparts such as GPT-2 \cite{radford2019language}. 

While LLMs are trained on giant text corpora, they  may not be representative of the specific domain of interest, in this case, speech transcriptions. This may result in limited generalization ability without domain-specific fine-tuning. Further, ASR applications warrant robustness to noise and other distortions, which text-only LLMs are incapable of handling on their own at rescoring time.

A potential solution to mitigate these limitations is to incorporate the speech input into LLM rescorers. Recent studies have demonstrated the effectiveness of leveraging audio information during second-pass rescoring \cite{sainath19_interspeech, audio_att_2020, hu2020deliberation, hu2022improving} to improve performance.  However, a tight integration of rescorer, attending to a shared speech encoder 
used in the first-pass, relies on ASR architecture, training mechanism and internal features, limiting the flexibility of being applied to other ASR systems.

Inspired by recent multi-modal LLM works~\cite{multimodal_Tsimpoukelli_NEURIPS2021, gao22e_interspeech, bapna2021slam, chen22r_interspeech}, we propose \mymethod, a multi-modal MLM rescorer, which is compatible with encapsulated ASR systems: our method by design can work with any first-pass ASR models (Hybrid / CTC / Transducer).
The rescorer is agnostic to ASR architecture, training mechanism and internal features, leading to better generalization capability. 
To the best of our knowledge, this is the first work to integrate a pre-trained self-supervised learning (SSL) speech representation model \cite{Baevski2019vqwav2vecSL,wav2vec2.0, hubert, Chen2021WavLMLS} into the second-pass rescoring. 
One key challenge of incorporating acoustic information into LLMs is to transform the speech into a form that can be accepted by the language model. 
We overcome this by using a cross-modal adaptation module consisting of Convolutional Neural Network (CNN) \cite{cnn} and adapter network \cite{houlsby_adapter}. 
We experiment with different auxiliary alignment losses for audio-text alignment, to effectively learn shared representations across the two modalities, and adopt contrastive learning which significantly improves the model performance.
Empirically, we show that MATE transfers well to new domains in zero-shot and few-shot settings, outperforming text-only baselines.  

\begin{figure}[ht]
\centering
\includegraphics[width=0.95\columnwidth]{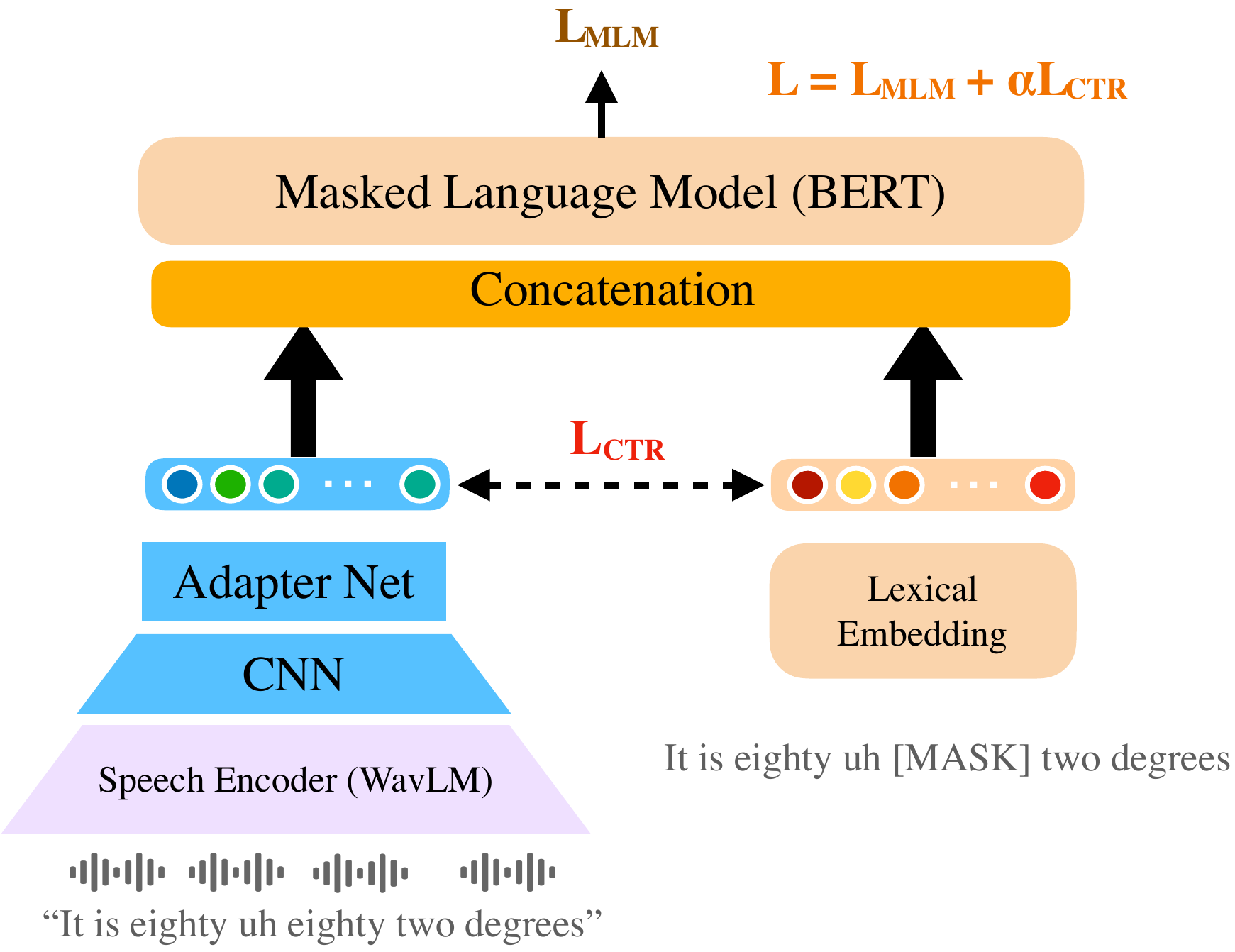}
\caption{\mymethod is trained with two losses: (1) The MLM which takes concatenated cross-modal representation as input and computes $\mathcal{L_{\tiny\textrm{MLM}}}$ on masked text tokens. (2)$\mathcal{L_{\tiny\textrm{CTR}}}$ to align the audio and text latent representations.
}
\label{fig: System Overview}
\vspace{-6mm}
\end{figure}


\section{Approach}


\mymethod consists of a pre-trained masked language model BERT, an self-supervised learning (SSL) based speech encoder WavLM \cite{Chen2021WavLMLS} and a modality matching module (CNN and adapter network), as illustrated in Figure~\ref{fig: System Overview}.

\subsection{System Architecture}
\paragraph{Masked Language Model}We use BERT, a pre-trained bidirectional MLM, as the primary component of our rescorer. In this work, we extend BERT to incorporate speech data along with text. The pre-trained embedding layers of BERT serve as the text embedding module, while the intermediate encoder layers take both acoustic and lexical representations as input.


\paragraph{Pre-trained Speech Encoder}To extract the acoustic representation, we use WavLM model, pre-trained on masked speech prediction and speech denoising tasks, achieving state-of-the-art performance on various speech processing tasks and outperforming other models like Wav2Vec2\cite{wav2vec2.0} and HuBERT\cite{hubert} on SUPERB \cite{superb} benchmark.



\paragraph{Cross-modal Adaptation} To align the acoustic and lexical representations in the same feature space, we design a cross-modal adaptation module. It is composed of two sub-modules: (i) Convolutional Neural Network (CNN) based subsampling component, to balance the sequence length between the modalities, and (ii) A bottleneck adapter network to project the acoustic representations to the BERT encoder input space. The outputs from the adapter network $a$ and lexical embedding $l$ are concatenated\footnote{We have also experimented with a cross-attention based merging mechanism, which leads to inferior performance.} horizontally $a ^\frown l$, and passed through the BERT encoder layers to fuse the information from the two modalities.

\subsection{Alignment Loss}
Pre-trained Masked Language Models are trained on text corpora~\cite{devlin-etal-2019-bert}. 
To explicitly align audio and text modalities, we propose introducing an explicit alignment loss function, thereby further enhancing the quality of cross-modal learning.


We adopt a contrastive loss function to enforce the mapping of acoustic representations $a$ and lexical representations $l$ to a shared feature space.
We conduct average pooling at utterance level, and denote the pooled vectors by $(\bar{a}_i, \bar{l}_j)$, from the acoustic or lexical representation $a_i$ and $l_i$ respectively.
Given acoustic-lexical representations $(\bar{a}_i, \bar{l}_i )_{1 \leq i \leq N}$ where $N$ is the batch size, we use the paired vectors $(\bar{a}_i, \bar{l}_i )$ as positive samples and the unpaired vectors $(\bar{a}_i, \bar{l}_j )_{i\neq j}$ in the same mini-batch as negative samples. The training objective is to minimize the following contrastive loss $\mathcal{L_{\tiny\textrm{CTR}}}$ with Negative Log-Likelihood (NLL) function:


\begin{equation}
 \mathcal{L_{\tiny\textrm{CTR}}}  = - \sum_{ i=1 }^N \textrm{log} \frac{\exp(\textrm{sim}(\bar{a}_i, \bar{l}_i))}     {\sum_{ j=1 }^N \exp(\textrm{sim}(\bar{a}_i, \bar{l}_j))}
\end{equation}

where $\textrm{sim}(\cdot, \cdot)$ is a similarity metric, implemented as dot product in our experiments.
Contrastive loss promotes a higher level of similarity between paired acoustic and lexical representations, as compared to unpaired representations, thus enhancing the alignment between the two modalities.

\subsection{Training and Inference}
\paragraph{Training} \mymethod is trained jointly
on the MLM objective $\mathcal{L_{\tiny\textrm{MLM}}}$, similar to that employed in the pre-training of BERT, and the contrastive loss $\mathcal{L_{\tiny\textrm{CTR}}}$. 

\begin{equation}
\label{eq:main}
 \mathcal{L} = \mathcal{L_{\tiny\textrm{MLM}}} + \alpha \cdot \mathcal{L_{\tiny\textrm{CTR}}}  
\end{equation}
Following BERT pre-training, a portion of tokens in the text sequence are randomly selected for prediction, and are replaced by the {\small[MASK]} token, a random token or left unchanged. 
In order to optimize the model's performance, the model is trained end-to-end and all the parameters are updated during the training process.

\paragraph{Inference} 
We use pseudo-log-likelihood (PLL) scoring \cite{wang-cho-2019-bert-pll, salazar-etal-2020-masked} to compute sequence level scores. Given an acoustic sequence $a=(s_1,..., s_{R})$ and a lexical sequence $l=(t_1,..., t_T)$, let $l_{\backslash k}=(t_1,..., t_{k-1}, {\small\textrm{ [MASK]}},  t_{k+1},  ..., t_T)$, PLL score is computed by summing conditional log probabilities $\textrm{log} P_{\tiny\textrm{MLM}} (l_i |a, l_{\backslash i} )$ of each masked lexical token:


\begin{equation}
\label{eq:pll}
\textrm{PLL}(l) = \sum_{ i=1 }^T \textrm{log} P_{\tiny\textrm{MLM}} (l_i |a, l_{\backslash i} )
\end{equation}
The final score of an utterance is computed as a linear interpolation of the first-pass ASR confidence score and second-pass PLL score, leveraging the complementary information to improve performance while allowing a trade-off between them.

\begin{table*}[t!]

\huge

\centering


\resizebox{\linewidth}{!}
{
\begin{tabular}{cc|cccccccc|cccccc}
\hline
\multicolumn{2}{l|}{\multirow{3}{*}{}} & \multicolumn{8}{c|}{\textbf{In-domain}}                                                                                                                                            & \multicolumn{6}{c}{\textbf{Out-of-domain}}                                                                  \\
\multicolumn{2}{l|}{}                  & \multicolumn{2}{c}{\textbf{MTDialogue}} & \multicolumn{2}{c}{\textbf{LS test-clean}} & \multicolumn{2}{c}{\textbf{LS test-other}} & \multicolumn{2}{c|}{\textbf{Voxpopuli}} & \multicolumn{2}{c}{\textbf{WSJ}} & \multicolumn{2}{c}{\textbf{ConvAI}} & \multicolumn{2}{c}{\textbf{SLURP}} \\
\multicolumn{2}{l|}{}                  & \textbf{WER}          & \textbf{CWER}          & \textbf{WER}         & \textbf{CWER}       & \textbf{WER}         & \textbf{CWER}       & \textbf{WER}       & \textbf{CWER}      & \textbf{WER}    & \textbf{CWER}  & \textbf{WER}     & \textbf{CWER}    & \textbf{WER}     & \textbf{CWER}   \\ \hline
(1)          & No rescoring            & 9.47                  & 14.63                  & 6.75                 & 8.07                & 11.98                & 15.61               & 11.06              & 10.33              & 8.16            & 8.75           & 5.89             & 9.00             & 24.91            & 29.53           \\ \hline
(2)          & GPT2-text              & 9.32                  & 14.37                  & 6.45                 & 7.78                & 11.70                & 15.11               & 10.72              & 9.94               & 7.64            & 8.40           & 5.76             & 8.66             & 24.91            & 29.53           \\
(3)          & BERT-text            & 9.05                  & 13.88                  & 5.50                 & 7.20                & 10.70                & 14.45               & 10.33              & 9.96               & 6.46            & 8.20           & 5.38             & 8.37             & 24.48            & 29.27           \\ \hline
(4)          & LAS rescoring           & 9.27                  & 14.15                  & 6.7                  & 7.99                & 11.97                & 15.59               & 11.02              & 10.21              & 8.01            & 8.55           & 5.81             & 8.84             & 24.91            & 29.53           \\

(5)          & Multi-modal-GPT2        & 9.24                  & 14.17                  & 6.35                 & 7.69                & 11.54                & 14.93               & 10.56              & 9.83               & 7.55            & 8.20           & 5.69             & 8.59             & 24.89            & 29.40           \\ 
\hline
(6)          & MATE-NA                 & 9.05                  & 13.90                  & 5.55                 & 7.29                & 10.75                & 14.51               & 10.34              & 9.92               & 6.49            & 8.10           & 5.40             & 8.36             & 24.46            & 29.24           \\
(7)          & MATE-MSE                & \textbf{7.49}         & \textbf{11.41}         & 5.22                 & 6.95                & 10.31                & 13.97               & 10.10              & 9.62               & 6.10            & 7.65           & \textbf{5.07}    & 7.92             & 23.84            & 28.24           \\
(8)         & MATE (\textit{ours})                    & 7.64                  & 11.70                  & \textbf{5.16}        & \textbf{6.84}       & \textbf{10.30}       & \textbf{13.81}      & \textbf{9.91}      & \textbf{9.47}      & \textbf{6.01}   & \textbf{7.46}  & 5.10             & \textbf{7.91}    & \textbf{23.77}   & \textbf{28.14}  \\ 
\hline


\multicolumn{16}{c}{\textit{Parameter-Efficient Tuning}} \\
\hline
(9)          & Frozen-ME           & 9.21                  & 14.22                  & 5.57                 & 7.34                & 10.82                & 14.65               & 10.37              & 9.80               & 6.55            & 8.15           & 5.42             & 8.34             & 24.39            & 29.13           \\
(10)          & WavLM-adapter           & 9.15                  & 14.02                  & 5.58                 & 7.41                & 10.81                & 14.69               & 10.23              & 9.86               & 6.52            & 8.05           & 5.47             & 8.39             & 24.56            & 29.27           \\
(11)          & ME-adapter             & 9.19                  & 14.12                  & 5.56                 & 7.43                & 10.79                & 14.63               & 10.09              & 9.60               & 6.43            & 8.20           & 5.42             & 8.35             & 24.34            & 29.08           \\ \hline
\end{tabular}

}


\caption{Performance measured by WER $\downarrow$ and CWER $\downarrow$. All models except (2-3) are multi-modal. 
\textit{(2) GPT2-text} \cite{gpt2}: Full fine-tuning of GPT2 on training corpora transcriptions.
\textit{(3) BERT-text} \cite{salazar-etal-2020-masked, devlin-etal-2019-bert}: Full fine-tuning of BERT on training corpora transcriptions, also denoted as "text-only baseline".
\textit{(4) LAS rescoring} \cite{sainath19_interspeech}: A multi-modal  baseline with LAS head rescoring (attention based LSTM decoder) accepting acoustic information from WavLM. 
\textit{(5) Multi-modal-GPT2:} A multi-modal uni-directional baseline with GPT2, accepting acoustic information from WavLM. 
\textit{(6) \mymethod-NA}: \mymethod without additional alignment loss; \textit{(7) \mymethod-MSE}: \mymethod trained with MSE loss instead of contrastive loss.  
\textit{(9) Frozen-ME (Masked Encoder)}: Fine-tune all parameters in multi-modal system except masked encoder (BERT) layers with only MLM objective. \textit{(10) WavLM-adapter}: add bottleneck adapter to speech encoder (WavLM) and do adapter-tuning on WavLM, all other parameters are frozen. \textit{(11) ME (Masked Encoder)-adapter}: do adapter-tuning on masked encoder (BERT), all other parameters are frozen.
}
\label{table_main}
\end{table*}

\section{Experiments
}



\subsection{Datasets}

\paragraph{Training Set} The training corpora consist of 10K+ hours of paired audio-text data, sampled from both public and in-house datasets. This data regime is representative of a variety of ASR systems used for various speech applications, with a mix of accents, speakers, sampling rates, and background noise. Less than 5\% of the data are synthetic audios generated using AWS Polly Text-to-Speech (TTS)~\footnote{https://aws.amazon.com/polly/} neural backend.

\paragraph{Evaluation Set} We evaluate the proposed \mymethod approach on both synthetic and real datasets from various domains: \textit{MTDialogue}
(movie-twitter), \textit{LibriSpeech (LS)}~\cite{Panayotov2015} and \textit{VoxPopuli}~\cite{Wang2021} are  in-domain sets, as the training set includes their corresponding train data splits. \textit{Wall Street Journal (WSJ)}~\cite{LDC93S6A}, \textit{ConvAI} (in-house), \textit{SLURP}~\cite{bastianelli-etal-2020-slurp} datasets are out-of-domain (OOD) datasets for zero-shot evaluation.



\textbf{MTDialogue}
(movie-twitter) is based on a public lexical dialogue corpus~\footnote{https://github.com/Phylliida/Dialogue-Datasets} which consists of movie subtitles and twitter user interactions. The audios are generated from TTS system. MTDialogue dataset is a seen dataset for open-book evaluation; i.e., all its data samples are covered in training data. An subset of 1.2 hour is sampled for evaluation. \textbf{LibriSpeech(LS)}~\cite{Panayotov2015} 
is a read English speech corpus based on LibriVox audiobooks. 
We consider the two official evaluation sets:  \textit{test-clean} and \textit{test-other}, each with 5.0 hours of test audios. \textbf{VoxPopuli}~\cite{Wang2021} 
consists of public political speech, sampled from 2009-2020 European Parliament event recordings. For our evaluation purpose, we utilize a 5-hour subset of VoxPopuli English data. 

We also evaluate \mymethod on OOD evaluation sets: ConvAI, WSJ, and SLURP.
The \textbf{Wall Street Journal (WSJ)}~\cite{LDC93S6A} corpus 
contains conventional and spontaneous dictation by journalists.
The \textit{test\_eval93} split of 0.4 hour is selected for our evaluation.
\textbf{ConvAI} 
is based on in-house user utterances of a task-oriented conversational AI system. The typical usage scenarios include booking flights, ordering food and querying health insurance information, etc. The 2.0 hours of audios are generated from TTS system. 
\textbf{SLURP}~\cite{bastianelli-etal-2020-slurp} 
is a public dataset for smart home virtual assistant development. Top usage scenarios include checking calendar, playing music, and asking about time, etc. We utilized the 10 hr test set for evaluation. 

\textbf{Ethical Considerations:} We have reviewed all licenses of public datasets, which allow the usage for research and paper publication. The in-house dataset ConvAI is internally approved for research purposes. All datasets are sets are de-identified to ensure anonymity. We also make sure the datasets cover various English accents, speakers and backgrounds.

\subsection{Evaluation Metrics}
We use word error rate (\textbf{WER}) and content word error rate (\textbf{CWER}) as the evaluation metrics.
CWER is computed on content words only (e.g., ``pizza'', ``parliament'', ``airline''), where we apply rule based method to filter out a predefined block-list of function words. 
Furthermore, we evaluate Spoken Language Understanding (\textbf{SLU}) performance on SLURP dataset using standard SLU metrics (accuracy and F1 score); SLU predictions (scenario, action and entity) are generated by a bi-directional Long Short-Term Memory (BiLSTM) NLU module (Appendix \ref{appendix_slurp}).



\section{Results and Analysis} 
We summarize the observations and analysis of the results from our experiments \footnote{Appendix~\ref{appendix_exp_setup} contains experimental setup details, including hyperparameters and infrastructure setting.} as follows:





\paragraph{\mymethod excels at both in-domain and out-of-domain generalization:} Table~\ref{table_main} summarizes the performance of the proposed \mymethod and multiple baseline models, under various settings, across in-domain and OOD datasets. Overall, we observe that our proposed approach (row 8) significantly outperforms text-only baseline (row 3) on in-domain datasets indicating that audio information helps even when we have sufficient target domain corpus for fine-tuning. Furthermore, results on OOD datasets indicate that \mymethod generalizes much better to new domains in the complete absence of domain data (zero-shot setting), when compared to the text-only baseline, by utilizing the rich information from audio.




\paragraph{MLMs are more effective multi-modal rescorers than autoregressive LMs:} Rows 2-5 indicate a significant performance gap between BERT and autoregressive rescorers (LAS/GPT-2). BERT-Text, which is a text-only baseline, outperforms even the multi-modal GPT2 indicating the root cause of the gap is the
lack of bi-directional (left and right) context in GPT2 which is necessary
for reliable and effective LLM scoring, hence validating the choice of MLM in \mymethod.



\paragraph{Alignment loss gives significant performance boost:} To study the effect of alignment loss, we train the multi-modal rescorer with two loss functions: Mean squared error (MSE) loss and contrastive loss.
Significant performance gains (row 6 vs. row 7-8) in Table~\ref{table_main} indicate that explicit alignment techniques greatly improve learning of multi-modal representations.
Specifically, contrastive loss not only aligns relevant pairs like MSE loss, but also promotes distancing irrelevant samples, leading to improved generalization on OOD sets.


\paragraph{Parameter-efficient fine-tuning results in limited gains:}

Rows 9-11 study the performance of a multi-modal rescorer under different parameter efficient fine-tuning settings. We observe that performance degrades as we move from full fine-tuning to adapter-tuning and freezing the full BERT encoder layers, indicating that fine-tuning BERT encoder is the most beneficial in terms of performance improvement. As expected, in comparison to model with full fine-tuning (row 6), rows 9-11 exhibit lower performance. This suggests that frozen or parameter-efficient training methods may lack the model capacity to fully leverage the acoustic information present in the multi-modal data.

\begin{figure}[ht]
\centering
\includegraphics[width=0.9\columnwidth]{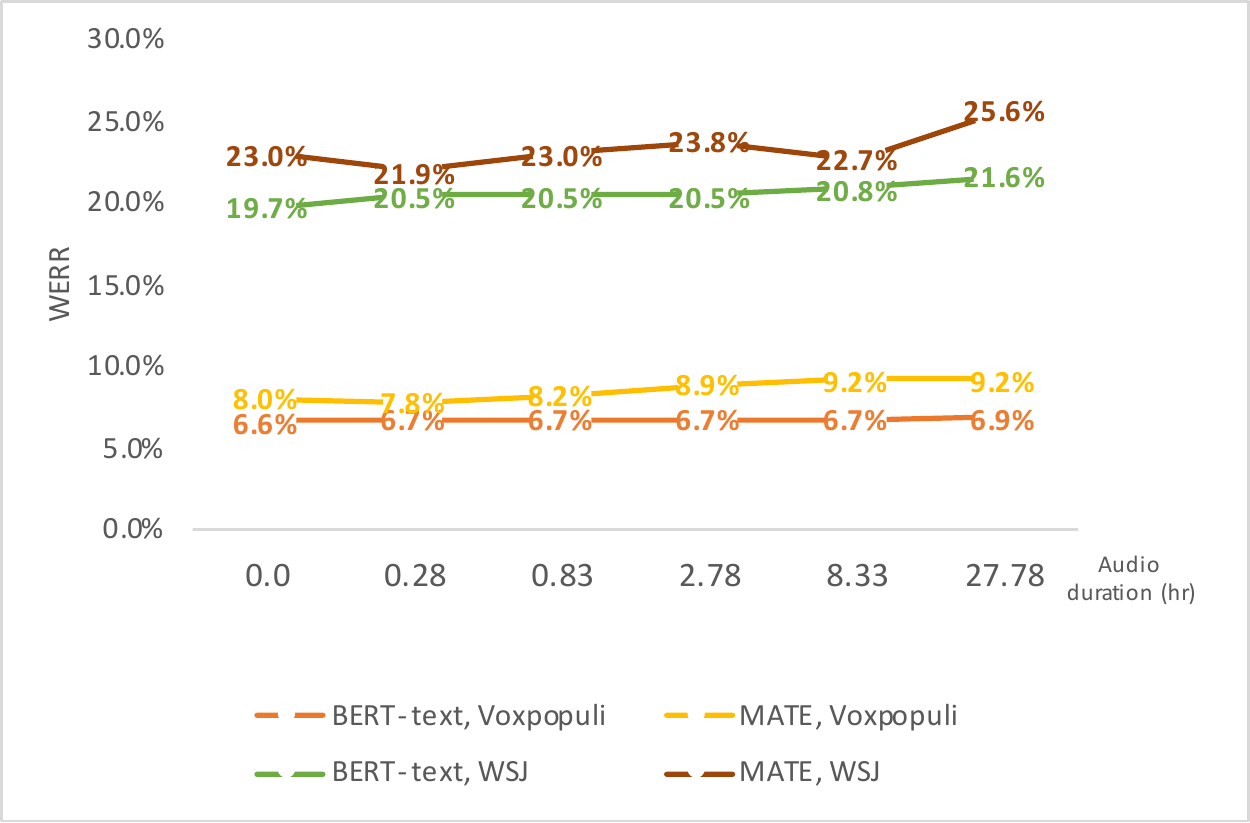}
\caption{Relative WER reduction (over first-pass) versus domain specific training data size.}
\label{fig: Fewshot}
\end{figure}

\paragraph{\mymethod is the most effective few-shot learner:} To study the effect of few-shot learning,
we plot the relative WER reduction (WERR) on Voxpopuli and WSJ datasets across different resource conditions as shown in Figure~\ref{fig: Fewshot}. We observe that \mymethod transfers well to the new domains in the zero-shot setting with no training or domain data at all. Few-shot performance clearly improves with more examples and goes a reasonable way towards closing the gap from zero-shot performance to full fine-tuning performance. We also observe that \mymethod consistently has superior performance to text-only baseline across both datasets, confirming the ability to rapidly adapt to new domains by leveraging additional information from the audio modality.


\begin{table}[t!]

\small

\centering

\begin{tabular}{lccc}
\hline
\multicolumn{1}{c}{\textbf{}} & \textbf{Scenario} & \textbf{Action} & \textbf{Entity} \\ \hline
No rescoring                  & 78.01             & 72.53           & 53.23           \\
GPT2-text                  & 78.01             & 72.53           & 53.23           \\
Multi-modal-GPT2               & 78.07             & 72.65           & 53.26           \\
BERT-text                 & 77.72             & 72.45           & 53.37           \\
\mymethod                     & \textbf{78.76}    & \textbf{73.70}  & \textbf{54.26}  \\ \hline
\end{tabular}
\caption{Zero-shot evaluation on SLURP SLU task: Accuracy for Scenario/Action, and F1 score for Entity. }
\label{table_slurp_slu}
\end{table}

\paragraph{\mymethod achieves best zero-shot performance improvement on downstream SLU tasks:} To evaluate the effectiveness of the proposed approach on the end goals in a dialog system, we compare it with other baselines using metrics such as scenario/action accuracy and entity F1 score in a zero-shot setting on SLURP dataset. From results
in Table~\ref{table_slurp_slu}, we observe that \mymethod consistently outperforms\footnote{The SLURP is a challenging corpus, which mimics the noisy use cases of smart home assistants. Hence, by improving rescoring method alone, we achieve less than 2\% absolute improvement in WER and SLU metrics.} the other baselines on end-to-end goals indicating that the improvements are mainly on
recognition of content words and slot entities.
\footnote{Qualitative examples are presented in Appendix \ref{appendix_qualitative}.}


\section{Conclusions}
We propose a novel multi-modal rescorer, \mymethod, which achieves significant WER, CWER reduction on  in-domain and OOD datasets. In zero-shot and few-shot settings, \mymethod performs well on unseen domains 
and adapts rapidly with limited data. The domain generalization capability of \mymethod makes it an effective choice as a second-pass rescorer for scaling ASR systems to new domains.

\section{Limitations}
One limitation of our approach is that incorporating acoustic features from an SSL speech encoder, in our case WavLM, introduces extra latency overhead, as we use a standalone ASR model for first-pass. Therefore, our approach may not be appropriate for certain applications that have exceptionally low latency constraints. 

Another limitation is that while multi-modal LLMs have the potential to improve ASR performance, they can be more complex and harder to interpret than text-only LLMs. This makes it more challenging to understand the model's decision making process or debug any potential errors.



\newpage

\nocite{*} 
\bibliography{anthology,custom}

\begin{thebibliography}{55}
\expandafter\ifx\csname natexlab\endcsname\relax\def\natexlab#1{#1}\fi

\bibitem[{Baevski et~al.(2019)Baevski, Schneider, and
  Auli}]{Baevski2019vqwav2vecSL}
Alexei Baevski, Steffen Schneider, and Michael Auli. 2019.
\newblock vq-wav2vec: Self-supervised learning of discrete speech
  representations.
\newblock \emph{ArXiv}, abs/1910.05453.

\bibitem[{Baevski et~al.(2020)Baevski, Zhou, Mohamed, and Auli}]{wav2vec2.0}
Alexei Baevski, Yuhao Zhou, Abdelrahman Mohamed, and Michael Auli. 2020.
\newblock \href
  {https://proceedings.neurips.cc/paper/2020/file/92d1e1eb1cd6f9fba3227870bb6d7f07-Paper.pdf}
  {wav2vec 2.0: A framework for self-supervised learning of speech
  representations}.
\newblock In \emph{Advances in Neural Information Processing Systems},
  volume~33, pages 12449--12460. Curran Associates, Inc.

\bibitem[{Bahdanau et~al.(2016)Bahdanau, Chorowski, Serdyuk, Brakel, and
  Bengio}]{e2e-att-lv}
Dzmitry Bahdanau, Jan Chorowski, Dmitriy Serdyuk, Philémon Brakel, and Yoshua
  Bengio. 2016.
\newblock \href {https://doi.org/10.1109/ICASSP.2016.7472618} {End-to-end
  attention-based large vocabulary speech recognition}.
\newblock In \emph{2016 IEEE International Conference on Acoustics, Speech and
  Signal Processing (ICASSP)}, pages 4945--4949.

\bibitem[{Bapna et~al.(2021)Bapna, Chung, Wu, Gulati, Jia, Clark, Johnson,
  Riesa, Conneau, and Zhang}]{bapna2021slam}
Ankur Bapna, Yu-an Chung, Nan Wu, Anmol Gulati, Ye~Jia, Jonathan~H Clark,
  Melvin Johnson, Jason Riesa, Alexis Conneau, and Yu~Zhang. 2021.
\newblock Slam: A unified encoder for speech and language modeling via
  speech-text joint pre-training.
\newblock \emph{arXiv preprint arXiv:2110.10329}.

\bibitem[{Bastianelli et~al.(2020)Bastianelli, Vanzo, Swietojanski, and
  Rieser}]{bastianelli-etal-2020-slurp}
Emanuele Bastianelli, Andrea Vanzo, Pawel Swietojanski, and Verena Rieser.
  2020.
\newblock \href {https://doi.org/10.18653/v1/2020.emnlp-main.588} {{SLURP}: A
  spoken language understanding resource package}.
\newblock In \emph{Proceedings of the 2020 Conference on Empirical Methods in
  Natural Language Processing (EMNLP)}, pages 7252--7262, Online. Association
  for Computational Linguistics.

\bibitem[{Brown et~al.(2020)Brown, Mann, Ryder, Subbiah, Kaplan, Dhariwal,
  Neelakantan, Shyam, Sastry, Askell, Agarwal, Herbert-Voss, Krueger, Henighan,
  Child, Ramesh, Ziegler, Wu, Winter, Hesse, Chen, Sigler, Litwin, Gray, Chess,
  Clark, Berner, McCandlish, Radford, Sutskever, and Amodei}]{gpt3}
Tom Brown, Benjamin Mann, Nick Ryder, Melanie Subbiah, Jared~D Kaplan, Prafulla
  Dhariwal, Arvind Neelakantan, Pranav Shyam, Girish Sastry, Amanda Askell,
  Sandhini Agarwal, Ariel Herbert-Voss, Gretchen Krueger, Tom Henighan, Rewon
  Child, Aditya Ramesh, Daniel Ziegler, Jeffrey Wu, Clemens Winter, Chris
  Hesse, Mark Chen, Eric Sigler, Mateusz Litwin, Scott Gray, Benjamin Chess,
  Jack Clark, Christopher Berner, Sam McCandlish, Alec Radford, Ilya Sutskever,
  and Dario Amodei. 2020.
\newblock \href
  {https://proceedings.neurips.cc/paper/2020/file/1457c0d6bfcb4967418bfb8ac142f64a-Paper.pdf}
  {Language models are few-shot learners}.
\newblock In \emph{Advances in Neural Information Processing Systems},
  volume~33, pages 1877--1901. Curran Associates, Inc.

\bibitem[{Cai et~al.(2023)Cai, Li, Jiang, Cho, Chen, Liu, Fan, and
  Guo}]{cai2023kgeco}
Jinglun Cai, Mingda Li, Ziyan Jiang, Eunah Cho, Zheng Chen, Yang Liu, Xing Fan,
  and Chenlei Guo. 2023.
\newblock \href {https://doi.org/10.1109/ICASSP49357.2023.10096826} {Kg-eco:
  Knowledge graph enhanced entity correction for query rewriting}.
\newblock In \emph{ICASSP 2023 - 2023 IEEE International Conference on
  Acoustics, Speech and Signal Processing (ICASSP)}.

\bibitem[{Chan et~al.(2016)Chan, Jaitly, Le, and Vinyals}]{LAS}
William Chan, Navdeep Jaitly, Quoc Le, and Oriol Vinyals. 2016.
\newblock \href {https://doi.org/10.1109/ICASSP.2016.7472621} {Listen, attend
  and spell: A neural network for large vocabulary conversational speech
  recognition}.
\newblock In \emph{2016 IEEE International Conference on Acoustics, Speech and
  Signal Processing (ICASSP)}, pages 4960--4964.

\bibitem[{Chen et~al.(2021)Chen, Wang, Chen, Wu, Liu, Chen, Li, Kanda,
  Yoshioka, Xiao, Wu, Zhou, Ren, Qian, Qian, Zeng, and Wei}]{Chen2021WavLMLS}
Sanyuan Chen, Chengyi Wang, Zhengyang Chen, Yu~Wu, Shujie Liu, Zhuo Chen, Jinyu
  Li, Naoyuki Kanda, Takuya Yoshioka, Xiong Xiao, Jian Wu, Long Zhou, Shuo Ren,
  Yanmin Qian, Yao Qian, Micheal Zeng, and Furu Wei. 2021.
\newblock Wavlm: Large-scale self-supervised pre-training for full stack speech
  processing.
\newblock \emph{IEEE Journal of Selected Topics in Signal Processing},
  16:1505--1518.

\bibitem[{Chen et~al.(2022)Chen, Zhang, Rosenberg, Ramabhadran, Moreno, Bapna,
  and Zen}]{chen22r_interspeech}
Zhehuai Chen, Yu~Zhang, Andrew Rosenberg, Bhuvana Ramabhadran, Pedro~J. Moreno,
  Ankur Bapna, and Heiga Zen. 2022.
\newblock \href {https://doi.org/10.21437/Interspeech.2022-10937} {{MAESTRO:
  Matched Speech Text Representations through Modality Matching}}.
\newblock In \emph{Proc. Interspeech 2022}, pages 4093--4097.

\bibitem[{Chi et~al.(2021)Chi, Salazar, and Kirchhoff}]{chi-etal-2021-align}
Ethan~A. Chi, Julian Salazar, and Katrin Kirchhoff. 2021.
\newblock \href {https://doi.org/10.18653/v1/2021.naacl-main.154}
  {Align-refine: Non-autoregressive speech recognition via iterative
  realignment}.
\newblock In \emph{Proceedings of the 2021 Conference of the North American
  Chapter of the Association for Computational Linguistics: Human Language
  Technologies}, pages 1920--1927, Online. Association for Computational
  Linguistics.

\bibitem[{Chorowski et~al.(2014)Chorowski, Bahdanau, Cho, and
  Bengio}]{e2e-att-fist-res}
Jan Chorowski, Dzmitry Bahdanau, Kyunghyun Cho, and Yoshua Bengio. 2014.
\newblock End-to-end continuous speech recognition using attention-based
  recurrent nn: First results.
\newblock In \emph{NIPS 2014 Workshop on Deep Learning, December 2014}.

\bibitem[{Delobelle et~al.(2022)Delobelle, Tokpo, Calders, and
  Berendt}]{delobelle2022measuring_fairness}
Pieter Delobelle, Ewoenam~Kwaku Tokpo, Toon Calders, and Bettina Berendt. 2022.
\newblock Measuring fairness with biased rulers: A comparative study on bias
  metrics for pre-trained language models.
\newblock In \emph{NAACL 2022: the 2022 Conference of the North American
  chapter of the Association for Computational Linguistics: human language
  technologies}, pages 1693--1706.

\bibitem[{Devlin et~al.(2019)Devlin, Chang, Lee, and
  Toutanova}]{devlin-etal-2019-bert}
Jacob Devlin, Ming-Wei Chang, Kenton Lee, and Kristina Toutanova. 2019.
\newblock \href {https://doi.org/10.18653/v1/N19-1423} {{BERT}: Pre-training of
  deep bidirectional transformers for language understanding}.
\newblock In \emph{Proceedings of the 2019 Conference of the North {A}merican
  Chapter of the Association for Computational Linguistics: Human Language
  Technologies, Volume 1 (Long and Short Papers)}, pages 4171--4186,
  Minneapolis, Minnesota. Association for Computational Linguistics.

\bibitem[{Gandhe and Rastrow(2020)}]{audio_att_2020}
Ankur Gandhe and Ariya Rastrow. 2020.
\newblock \href
  {https://www.amazon.science/publications/audio-attention-discriminative-language-model-for-asr-rescoring}
  {Audio-attention discriminative language model for asr rescoring}.
\newblock In \emph{ICASSP 2020}.

\bibitem[{Gao et~al.(2022)Gao, Ni, Qian, Zhang, Chang, and
  Hasegawa-Johnson}]{gao22e_interspeech}
Heting Gao, Junrui Ni, Kaizhi Qian, Yang Zhang, Shiyu Chang, and Mark
  Hasegawa-Johnson. 2022.
\newblock \href {https://doi.org/10.21437/Interspeech.2022-11031} {{WavPrompt:
  Towards Few-Shot Spoken Language Understanding with Frozen Language Models}}.
\newblock In \emph{Proc. Interspeech 2022}, pages 2738--2742.

\bibitem[{Garofolo et~al.(1993)Garofolo, Graff, Paul, and Pallett}]{LDC93S6A}
John~S. Garofolo, David Graff, Doug Paul, and David Pallett. 1993.
\newblock \href {https://catalog.ldc.upenn.edu/LDC93S6A} {{CSR-I} ({WSJ}0)
  {C}omplete {LDC93S6A}}.
\newblock \emph{Linguistic Data Consortium}.

\bibitem[{Graves(2012)}]{RNNT}
Alex Graves. 2012.
\newblock Sequence transduction with recurrent neural networks.
\newblock \emph{ArXiv}, abs/1211.3711.

\bibitem[{Graves and Jaitly(2014)}]{pmlr-v32-graves14}
Alex Graves and Navdeep Jaitly. 2014.
\newblock \href {https://proceedings.mlr.press/v32/graves14.html} {Towards
  end-to-end speech recognition with recurrent neural networks}.
\newblock In \emph{Proceedings of the 31st International Conference on Machine
  Learning}, volume~32 of \emph{Proceedings of Machine Learning Research},
  pages 1764--1772. PMLR.

\bibitem[{Gulati et~al.(2020)Gulati, Qin, Chiu, Parmar, Zhang, Yu, Han, Wang,
  Zhang, Wu, and Pang}]{Gulati2020}
Anmol Gulati, James Qin, Chung~Cheng Chiu, Niki Parmar, Yu~Zhang, Jiahui Yu,
  Wei Han, Shibo Wang, Zhengdong Zhang, Yonghui Wu, and Ruoming Pang. 2020.
\newblock \href {https://doi.org/10.21437/Interspeech.2020-3015} {Conformer:
  Convolution-augmented transformer for speech recognition}.
\newblock In \emph{Proceedings of the Annual Conference of the International
  Speech Communication Association, INTERSPEECH}.

\bibitem[{Guo et~al.(2021)Guo, Rush, and Kim}]{guo-etal-2021-parameter}
Demi Guo, Alexander Rush, and Yoon Kim. 2021.
\newblock \href {https://doi.org/10.18653/v1/2021.acl-long.378}
  {Parameter-efficient transfer learning with diff pruning}.
\newblock In \emph{Proceedings of the 59th Annual Meeting of the Association
  for Computational Linguistics and the 11th International Joint Conference on
  Natural Language Processing (Volume 1: Long Papers)}, pages 4884--4896,
  Online. Association for Computational Linguistics.

\bibitem[{Hannun et~al.(2014)Hannun, Case, Casper, Catanzaro, Diamos, Elsen,
  Prenger, Satheesh, Sengupta, Coates, and Ng}]{deepspeech}
Awni~Y. Hannun, Carl Case, Jared Casper, Bryan Catanzaro, Greg Diamos, Erich
  Elsen, Ryan Prenger, Sanjeev Satheesh, Shubho Sengupta, Adam Coates, and
  Andrew~Y. Ng. 2014.
\newblock \href {http://arxiv.org/abs/1412.5567} {Deep speech: Scaling up
  end-to-end speech recognition}.
\newblock \emph{CoRR}, abs/1412.5567.

\bibitem[{Hinton et~al.(2012)Hinton, Deng, Yu, Dahl, rahman Mohamed, Jaitly,
  Senior, Vanhoucke, Nguyen, Sainath, and Kingsbury}]{hybrid-dnn}
Geoffrey Hinton, Li~Deng, Dong Yu, George Dahl, Abdel rahman Mohamed, Navdeep
  Jaitly, Andrew Senior, Vincent Vanhoucke, Patrick Nguyen, Tara Sainath, and
  Brian Kingsbury. 2012.
\newblock Deep neural networks for acoustic modeling in speech recognition.
\newblock \emph{Signal Processing Magazine}.

\bibitem[{Houlsby et~al.(2019)Houlsby, Giurgiu, Jastrzebski, Morrone,
  de~Laroussilhe, Gesmundo, Attariyan, and Gelly}]{houlsby_adapter}
Neil Houlsby, Andrei Giurgiu, Stanislaw Jastrzebski, Bruna Morrone, Quentin
  de~Laroussilhe, Andrea Gesmundo, Mona Attariyan, and Sylvain Gelly. 2019.
\newblock \href
  {http://dblp.uni-trier.de/db/conf/icml/icml2019.html#HoulsbyGJMLGAG19}
  {Parameter-efficient transfer learning for nlp.}
\newblock In \emph{ICML}, volume~97 of \emph{Proceedings of Machine Learning
  Research}, pages 2790--2799. PMLR.

\bibitem[{Hsu et~al.(2021)Hsu, Bolte, Tsai, Lakhotia, Salakhutdinov, and
  Mohamed}]{hubert}
Wei-Ning Hsu, Benjamin Bolte, Yao-Hung~Hubert Tsai, Kushal Lakhotia, Ruslan
  Salakhutdinov, and Abdelrahman Mohamed. 2021.
\newblock \href {https://doi.org/10.1109/TASLP.2021.3122291} {Hubert:
  Self-supervised speech representation learning by masked prediction of hidden
  units}.
\newblock \emph{IEEE/ACM Transactions on Audio, Speech, and Language
  Processing}, 29:3451--3460.

\bibitem[{Hu et~al.(2021)Hu, Pang, Sainath, and
  Strohman}]{transformer-deliberation-2021}
Ke~Hu, Ruoming Pang, Tara~N. Sainath, and Trevor Strohman. 2021.
\newblock \href {https://doi.org/10.1109/SLT48900.2021.9383497} {Transformer
  based deliberation for two-pass speech recognition}.
\newblock In \emph{2021 IEEE Spoken Language Technology Workshop (SLT)}, pages
  68--74.

\bibitem[{Hu et~al.(2022)Hu, Sainath, He, Prabhavalkar, Strohman, Mavandadi,
  and Wang}]{hu2022improving}
Ke~Hu, Tara~N Sainath, Yanzhang He, Rohit Prabhavalkar, Trevor Strohman, Sepand
  Mavandadi, and Weiran Wang. 2022.
\newblock Improving deliberation by text-only and semi-supervised training.
\newblock \emph{arXiv preprint arXiv:2206.14716}.

\bibitem[{Hu et~al.(2020)Hu, Sainath, Pang, and
  Prabhavalkar}]{hu2020deliberation}
Ke~Hu, Tara~N Sainath, Ruoming Pang, and Rohit Prabhavalkar. 2020.
\newblock Deliberation model based two-pass end-to-end speech recognition.
\newblock In \emph{ICASSP 2020-2020 IEEE International Conference on Acoustics,
  Speech and Signal Processing (ICASSP)}, pages 7799--7803. IEEE.

\bibitem[{Kingma and Ba(2014)}]{Kingma2014AdamAM}
Diederik~P. Kingma and Jimmy Ba. 2014.
\newblock Adam: A method for stochastic optimization.
\newblock \emph{CoRR}, abs/1412.6980.

\bibitem[{LeCun et~al.(1989)LeCun, Boser, Denker, Henderson, Howard, Hubbard,
  and Jackel}]{cnn}
Y.~LeCun, B.~Boser, J.~S. Denker, D.~Henderson, R.~E. Howard, W.~Hubbard, and
  L.~D. Jackel. 1989.
\newblock \href {https://doi.org/10.1162/neco.1989.1.4.541} {Backpropagation
  applied to handwritten zip code recognition}.
\newblock \emph{Neural Computation}, 1(4):541--551.

\bibitem[{Leng et~al.(2021)Leng, Tan, Zhu, Xu, Luo, Liu, Qin, Li, Lin, and
  Liu}]{fastcorrect_NEURIPS2021}
Yichong Leng, Xu~Tan, Linchen Zhu, Jin Xu, Renqian Luo, Linquan Liu, Tao Qin,
  Xiangyang Li, Edward Lin, and Tie-Yan Liu. 2021.
\newblock \href
  {https://proceedings.neurips.cc/paper_files/paper/2021/file/b597460c506e8e35fb0cc1c1905dd3bc-Paper.pdf}
  {Fastcorrect: Fast error correction with edit alignment for automatic speech
  recognition}.
\newblock In \emph{Advances in Neural Information Processing Systems},
  volume~34, pages 21708--21719. Curran Associates, Inc.

\bibitem[{McDermott et~al.(2019)McDermott, Sak, and
  Variani}]{density_ratio_19_lm_fusion}
Erik McDermott, Hasim Sak, and Ehsan Variani. 2019.
\newblock \href {https://doi.org/10.1109/ASRU46091.2019.9003790} {A density
  ratio approach to language model fusion in end-to-end automatic speech
  recognition}.
\newblock In \emph{2019 IEEE Automatic Speech Recognition and Understanding
  Workshop (ASRU)}, pages 434--441.

\bibitem[{Miao et~al.(2015)Miao, Gowayyed, and Metze}]{Miao2015EESENES}
Yajie Miao, Mohammad~Abdelaziz Gowayyed, and Florian Metze. 2015.
\newblock Eesen: End-to-end speech recognition using deep rnn models and
  wfst-based decoding.
\newblock \emph{2015 IEEE Workshop on Automatic Speech Recognition and
  Understanding (ASRU)}, pages 167--174.

\bibitem[{Panayotov et~al.(2015)Panayotov, Chen, Povey, and
  Khudanpur}]{Panayotov2015}
Vassil Panayotov, Guoguo Chen, Daniel Povey, and Sanjeev Khudanpur. 2015.
\newblock \href {https://doi.org/10.1109/ICASSP.2015.7178964} {Librispeech: An
  {ASR} corpus based on public domain audio books}.
\newblock In \emph{IEEE International Conference on Acoustics, Speech and
  Signal Processing (ICASSP)}.

\bibitem[{Radford et~al.(2019{\natexlab{a}})Radford, Wu, Child, Luan, Amodei,
  and Sutskever}]{gpt2}
A.~Radford, Jeffrey Wu, R.~Child, David Luan, Dario Amodei, and Ilya Sutskever.
  2019{\natexlab{a}}.
\newblock Language models are unsupervised multitask learners.

\bibitem[{Radford and Narasimhan(2018)}]{gpt1}
Alec Radford and Karthik Narasimhan. 2018.
\newblock Improving language understanding by generative pre-training.

\bibitem[{Radford et~al.(2019{\natexlab{b}})Radford, Wu, Child, Luan, Amodei,
  Sutskever et~al.}]{radford2019language}
Alec Radford, Jeffrey Wu, Rewon Child, David Luan, Dario Amodei, Ilya
  Sutskever, et~al. 2019{\natexlab{b}}.
\newblock Language models are unsupervised multitask learners.
\newblock \emph{OpenAI blog}, 1(8):9.

\bibitem[{Rekabsaz et~al.(2021)Rekabsaz, Kopeinik, and
  Schedl}]{societal_biases_21}
Navid Rekabsaz, Simone Kopeinik, and Markus Schedl. 2021.
\newblock \href {https://doi.org/10.1145/3404835.3462949} {Societal biases in
  retrieved contents: Measurement framework and adversarial mitigation of bert
  rankers}.
\newblock In \emph{Proceedings of the 44th International ACM SIGIR Conference
  on Research and Development in Information Retrieval}, SIGIR '21, page
  306–316, New York, NY, USA. Association for Computing Machinery.

\bibitem[{Sainath et~al.(2019)Sainath, Pang, Rybach, He, Prabhavalkar, Li,
  Visontai, Liang, Strohman, Wu, McGraw, and Chiu}]{sainath19_interspeech}
Tara~N. Sainath, Ruoming Pang, David Rybach, Yanzhang He, Rohit Prabhavalkar,
  Wei Li, Mirkó Visontai, Qiao Liang, Trevor Strohman, Yonghui Wu, Ian McGraw,
  and Chung-Cheng Chiu. 2019.
\newblock \href {https://doi.org/10.21437/Interspeech.2019-1341} {{Two-Pass
  End-to-End Speech Recognition}}.
\newblock In \emph{Proc. Interspeech 2019}, pages 2773--2777.

\bibitem[{Salazar et~al.(2020)Salazar, Liang, Nguyen, and
  Kirchhoff}]{salazar-etal-2020-masked}
Julian Salazar, Davis Liang, Toan~Q. Nguyen, and Katrin Kirchhoff. 2020.
\newblock \href {https://doi.org/10.18653/v1/2020.acl-main.240} {Masked
  language model scoring}.
\newblock In \emph{Proceedings of the 58th Annual Meeting of the Association
  for Computational Linguistics}, pages 2699--2712, Online. Association for
  Computational Linguistics.

\bibitem[{Schwartz and Austin(1991)}]{Schwartz1991nbest}
R.~Schwartz and Steve Austin. 1991.
\newblock A comparison of several approximate algorithms for finding multiple
  (n-best) sentence hypotheses.
\newblock \emph{ICASSP 91: 1991 International Conference on Acoustics, Speech,
  and Signal Processing}, pages 701--704 vol. 1.

\bibitem[{Sriram et~al.(2018)Sriram, Jun, Satheesh, and
  Coates}]{sriram18_code_fusion}
Anuroop Sriram, Heewoo Jun, Sanjeev Satheesh, and Adam Coates. 2018.
\newblock \href {https://doi.org/10.21437/Interspeech.2018-1392} {{Cold Fusion:
  Training Seq2Seq Models Together with Language Models}}.
\newblock In \emph{Proc. Interspeech 2018}, pages 387--391.

\bibitem[{Sung et~al.(2019)Sung, Liu, Lee, and Lee}]{Sung2019}
Tzu-Wei Sung, Jun-You Liu, Hung-yi Lee, and Lin-shan Lee. 2019.
\newblock \href {https://doi.org/10.1109/ICASSP.2019.8682801} {Towards
  end-to-end speech-to-text translation with two-pass decoding}.
\newblock In \emph{ICASSP 2019 - 2019 IEEE International Conference on
  Acoustics, Speech and Signal Processing (ICASSP)}, pages 7175--7179.

\bibitem[{Trentin and Gori(2001)}]{hybrid-survey}
Edmondo Trentin and Marco Gori. 2001.
\newblock \href {https://doi.org/https://doi.org/10.1016/S0925-2312(00)00308-8}
  {A survey of hybrid ann/hmm models for automatic speech recognition}.
\newblock \emph{Neurocomputing}, 37(1):91--126.

\bibitem[{Tsimpoukelli et~al.(2021)Tsimpoukelli, Menick, Cabi, Eslami, Vinyals,
  and Hill}]{multimodal_Tsimpoukelli_NEURIPS2021}
Maria Tsimpoukelli, Jacob~L Menick, Serkan Cabi, S.~M.~Ali Eslami, Oriol
  Vinyals, and Felix Hill. 2021.
\newblock \href
  {https://proceedings.neurips.cc/paper/2021/file/01b7575c38dac42f3cfb7d500438b875-Paper.pdf}
  {Multimodal few-shot learning with frozen language models}.
\newblock In \emph{Advances in Neural Information Processing Systems},
  volume~34, pages 200--212. Curran Associates, Inc.

\bibitem[{Udagawa et~al.(2022)Udagawa, Suzuki, Kurata, Itoh, and
  Saon}]{llm_22interspeech}
Takuma Udagawa, Masayuki Suzuki, Gakuto Kurata, Nobuyasu Itoh, and George Saon.
  2022.
\newblock \href {https://doi.org/10.21437/Interspeech.2022-11123} {{Effect and
  Analysis of Large-scale Language Model Rescoring on Competitive ASR
  Systems}}.
\newblock In \emph{Proc. Interspeech 2022}, pages 3919--3923.

\bibitem[{Vaswani et~al.(2017)Vaswani, Shazeer, Parmar, Uszkoreit, Jones,
  Gomez, Kaiser, and Polosukhin}]{transformer}
Ashish Vaswani, Noam Shazeer, Niki Parmar, Jakob Uszkoreit, Llion Jones,
  Aidan~N Gomez, \L~ukasz Kaiser, and Illia Polosukhin. 2017.
\newblock \href
  {https://proceedings.neurips.cc/paper/2017/file/3f5ee243547dee91fbd053c1c4a845aa-Paper.pdf}
  {Attention is all you need}.
\newblock In \emph{Advances in Neural Information Processing Systems},
  volume~30. Curran Associates, Inc.

\bibitem[{Wang and Cho(2019)}]{wang-cho-2019-bert-pll}
Alex Wang and Kyunghyun Cho. 2019.
\newblock \href {https://doi.org/10.18653/v1/W19-2304} {{BERT} has a mouth, and
  it must speak: {BERT} as a {M}arkov random field language model}.
\newblock In \emph{Proceedings of the Workshop on Methods for Optimizing and
  Evaluating Neural Language Generation}, pages 30--36, Minneapolis, Minnesota.
  Association for Computational Linguistics.

\bibitem[{Wang et~al.(2021)Wang, Rivière, Lee, Wu, Talnikar, Haziza,
  Williamson, Pino, and Dupoux}]{Wang2021}
Changhan Wang, Morgane Rivière, Ann Lee, Anne Wu, Chaitanya Talnikar, Daniel
  Haziza, Mary Williamson, Juan Pino, and Emmanuel Dupoux. 2021.
\newblock \href {https://doi.org/10.18653/V1/2021.ACL-LONG.80} {{V}ox{P}opuli:
  A large-scale multilingual speech corpus for representation learning,
  semi-supervised learning and interpretation}.
\newblock In \emph{Proceedings of the 59th Annual Meeting of the Association
  for Computational Linguistics and the 11th International Joint Conference on
  Natural Language Processing}.

\bibitem[{Xia et~al.(2017)Xia, Tian, Wu, Lin, Qin, Yu, and
  Liu}]{delibration_neurips17}
Yingce Xia, Fei Tian, Lijun Wu, Jianxin Lin, Tao Qin, Nenghai Yu, and Tie-Yan
  Liu. 2017.
\newblock \href
  {https://proceedings.neurips.cc/paper/2017/file/c6036a69be21cb660499b75718a3ef24-Paper.pdf}
  {Deliberation networks: Sequence generation beyond one-pass decoding}.
\newblock In \emph{Advances in Neural Information Processing Systems},
  volume~30. Curran Associates, Inc.

\bibitem[{Xu et~al.(2022)Xu, Gu, Kolehmainen, Khan, Gandhe, Rastrow, Stolcke,
  and Bulyko}]{rescoreBERT}
Liyan Xu, Yile Gu, Jari Kolehmainen, Haidar Khan, Ankur Gandhe, Ariya Rastrow,
  Andreas Stolcke, and Ivan Bulyko. 2022.
\newblock \href {https://doi.org/10.1109/ICASSP43922.2022.9747118}
  {Rescorebert: Discriminative speech recognition rescoring with bert}.
\newblock In \emph{ICASSP 2022 - 2022 IEEE International Conference on
  Acoustics, Speech and Signal Processing (ICASSP)}, pages 6117--6121.

\bibitem[{Yang et~al.(2021)Yang, Chi, Chuang, Lai, Lakhotia, Lin, Liu, Shi,
  Chang, Lin, Huang, Tseng, tik Lee, Liu, Huang, Dong, Li, Watanabe, Mohamed,
  and yi~Lee}]{superb}
Shu-Wen Yang, Po-Han Chi, Yung-Sung Chuang, Cheng-I~Jeff Lai, Kushal Lakhotia,
  Yist~Y. Lin, Andy~T. Liu, Jiatong Shi, Xuankai Chang, Guan-Ting Lin,
  Tzu-Hsien Huang, Wei-Cheng Tseng, Ko~tik Lee, Da-Rong Liu, Zili Huang, Shuyan
  Dong, Shang-Wen Li, Shinji Watanabe, Abdelrahman Mohamed, and Hung yi~Lee.
  2021.
\newblock \href {https://doi.org/10.21437/Interspeech.2021-1775} {{SUPERB:
  Speech Processing Universal PERformance Benchmark}}.
\newblock In \emph{Proc. Interspeech 2021}, pages 1194--1198.

\bibitem[{Zhang et~al.(2020)Zhang, Lu, Sak, Tripathi, McDermott, Koo, and
  Kumar}]{transformer-transducer}
Qian Zhang, Han Lu, Hasim Sak, Anshuman Tripathi, Erik McDermott, Stephen Koo,
  and Shankar Kumar. 2020.
\newblock Transformer transducer: A streamable speech recognition model with
  transformer encoders and rnn-t loss.
\newblock \emph{ICASSP 2020 - 2020 IEEE International Conference on Acoustics,
  Speech and Signal Processing (ICASSP)}, pages 7829--7833.

\bibitem[{Zhao et~al.(2019)Zhao, Sainath, Rybach, Rondon, Bhatia, Li, and
  Pang}]{zhao19d_shallow_fusion}
Ding Zhao, Tara~N. Sainath, David Rybach, Pat Rondon, Deepti Bhatia, Bo~Li, and
  Ruoming Pang. 2019.
\newblock \href {https://doi.org/10.21437/Interspeech.2019-1209}
  {{Shallow-Fusion End-to-End Contextual Biasing}}.
\newblock In \emph{Proc. Interspeech 2019}, pages 1418--1422.

\bibitem[{Zheng et~al.(2021)Zheng, Xiao, Gong, Zhou, Liang, and
  Lin}]{zheng-etal-2021-wav-bert}
Guolin Zheng, Yubei Xiao, Ke~Gong, Pan Zhou, Xiaodan Liang, and Liang Lin.
  2021.
\newblock \href {https://doi.org/10.18653/v1/2021.findings-emnlp.236}
  {Wav-{BERT}: Cooperative acoustic and linguistic representation learning for
  low-resource speech recognition}.
\newblock In \emph{Findings of the Association for Computational Linguistics:
  EMNLP 2021}, pages 2765--2777, Punta Cana, Dominican Republic. Association
  for Computational Linguistics.

\end{thebibliography}
\bibliographystyle{acl_natbib}

\clearpage
\section*{Appendix}
\appendix


\section{SLURP SLU semantics and NLU module}
\label{appendix_slurp}
SLURP dataset consists of user interactions with smart home virtual assistants. The semantics are annotated with three levels of semantics:
Scenario, Action and Entity. For example, ASR transcript ``how do I make a turkey'' is the annotated with semantics ``scenario: cooking $|$ action: recipe $|$ entities: [(type: food $|$ filler: turkey)]''. 
The SLU semantics spans over 18 different scenarios, 46 defined
actions and 55 different entity types \cite{bastianelli-etal-2020-slurp}. 

In the NLU module, we treat semantics prediction as a sequence-to-sequence problem. Specifically, given an ASR transcript after rescoring ``how do I make a turkey'', the goal is to predict: ``scenario: cooking $|$ action: recipe $|$ entities: [(type: food $|$ filler: turkey)]''. The NLU module has an encoder-decoder structure based on bi-directional Long Short-Term Memory (Bi-LSTM). Both the encoder and the decoder have hidden dimention 256. The encoder has 2 layers while the decoder has 3 layers. We use Negative Log-Likelihood  (NLL) loss for as training objective for sequence prediction. We train the model on ground truth <transcript, NLU semantics> paris from SLURP training dataset. The learning rate is set to 3e-4 and the training is conducted for 20 epochs with batch size 16. 


\section{Experimental Setup}
\label{appendix_exp_setup}
\mymethod has $217$M parameters in total. For both masked language model and speech encoder, we utilize base size models for efficiency (BERT-Base $110$M and WavLM-Base+ $95$M respectively). The convolutional network contains $3$ layers with $768$ channels with strides
$(2,1,2)$ and kernel widths $(3,1,1)$. The bottleneck adapter layer has compression factor $0.5$.  

The training experiment for \mymethod is conducted end-to-end: we train all modules simultaneously. We use Adam optimizer \cite{Kingma2014AdamAM} with linear decay of learning rate. We set initial learning rate to $5e-5$ and batch size to $32$. We searched the hyperparameter $\alpha$ in Eq.\ref{eq:main} with $(1.0, 3.0, 10.0)$, and the final value is set to $1.0$. The training was conducted for $88K$ steps. All the experiments are performed with NVIDIA Tesla V100 GPUs in a single run. The training for \mymethod takes 39.7 hours on a Tesla V100 8-GPU machine. 

Our first-pass ASR model has a conformer-CTC~\cite{Gulati2020} architecture. which is trained on $50$K+ hours audio-transcript paired data. The conformer encoder consists of $20$ layers of conformer blocks with hidden dimension $2048$; while the shallow decoder is a single Transformer-based layer with the same hidden dimension of $2048$. The conformer-CTC model has approximately $140$M parameters, 

We use SCTK\footnote{https://github.com/chinshr/sctk} package for WER and CWER evaluation. CWER has the same logic as WER computation except that we filter out function words. We use SLURP toolkit\footnote{https://github.com/pswietojanski/slurp} for SLU semantics evaluation.


\section{Attention Visualization}
\begin{figure}[ht]
\centering
\includegraphics[width=\columnwidth]{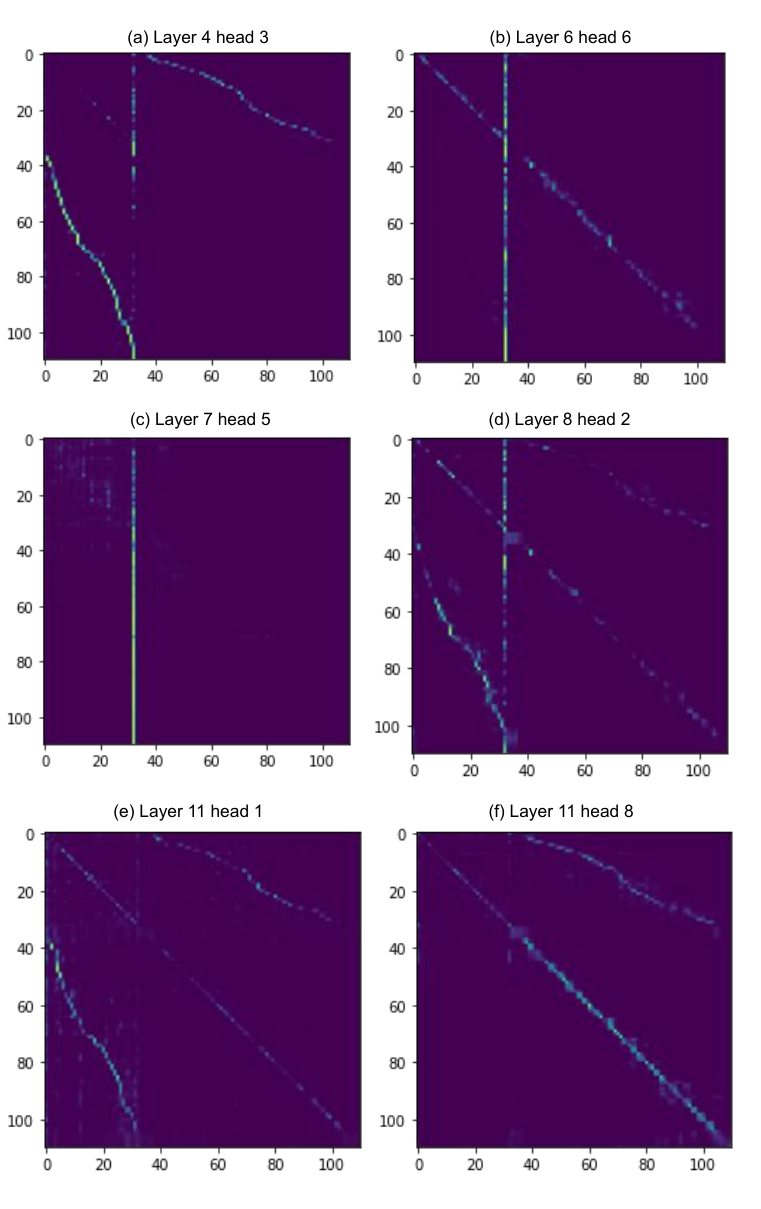}
\caption{Selected attention plots from the self-attention layers of the 12-layer BERT encoder The sample utterance (from wsj\_eval93) contains 110 total frames: the first 33 frames are lexical embedding, followed by 77 acoustic embedding frames. The utterance is: "last year new hampshire enacted legislation enabling banks from outside the state to acquire new hampshire banks but restrictions in the bill discouraged potential buyers"}
\label{fig: attention plot}
\end{figure}
We visualize the learned self-attention plots extracted from the proposed MATE model in Figure~\ref{fig: attention plot}.
The model has 12 Transformer layers and with 12 heads in each multi-head self-attention. We selected 6 representative plot from the 144 total attention plots with a sample utterance from wsj\_eval93 test set. 
The input utterance has 33 tokens and 77 frames for the acoustic feature, the acoustic features are appended to the lexical embedding before fed into the BERT model. Our observations are listed as follows:
\begin{itemize}
    \item \textbf{(a) (b) (c) and (d)} The plots highlight the border of the text input and audio input (the vertical straight line on position 32). We can conclude that even without feeding any modality border information to \mymethod, it can learn the border of two modalities itself.
    \item \textbf{(a), (d), (e) and (f)} The monotonic audio-to-text position alignment is clearly shown in the plots. This indicates that the acoustic and lexical representations are successfully mapped to one unified feature space. Interestingly, plots (a), (e) and (f) show that text-to-audio position alignment can also be learned by \mymethod.
\end{itemize}



\section{Risks}
The proposed system, \mymethod, incorporates both pre-trained language model (BERT) and speech model (WavLM) into its design. Such pre-trained models can contain biases and stereotypes against certain religion, race and gender groups~\cite{societal_biases_21,delobelle2022measuring_fairness}.

\section{Qualitative Examples}
\label{appendix_qualitative}
To further understand why the proposed approach, \mymethod yields more accurate prediction, we selected several representative cases from the evaluation sets. Table~\ref{tab:case} clearly shows that \mymethod tends to correct more vocabulary or grammar errors present in the n-best list. We observe \mymethod is able to correct many ASR errors which are not resolvable by text information alone. In the example from SLURP, both ``who in the hallway'' and ``hoover the hallway'' are plausible utterances in an informal style of daily speech. With the aid of acoustic information, \mymethod is able to assign higher score to the correct utterance ``hoover the hallway''.

\begin{table*}[t]
\resizebox{\linewidth}{!}
{
\begin{tabular}{lll}
\hline
\textbf{Dataset} & \textbf{}                      & \textbf{Utterance}                                                                         \\ \hline
SLURP            & Ground Truth                   & remove tuesday alarm of   nine a m                                                         \\
                 & Rescored   1-best by BERT-text & \red{move to} alarm of nine a m                                                                  \\
                 & Rescored   1-best by MATE      & \blue{remove tuesday} alarm at nine a m                                                           \\ \cline{2-3} 
                 & Ground Truth                   & hoover the hallway                                                                         \\ 
                 & Rescored   1-best by BERT-text & \red{who} in the hallway                                                                         \\
                 & Rescored   1-best by MATE      & \blue{hoover} the hallway     \\ \cline{2-3}                        
                 & Ground Truth                   & cancel business meeting on wednesday                                                         \\
                 & Rescored   1-best by BERT-text &
                 \red{council} business meeting on wednesday                                                        \\
                 & Rescored   1-best by MATE      & \blue{cancel} business meeting on wednesday                                                         \\ \cline{2-3}
                 & Ground Truth                   & can you let {delta} know i am never using them again                                                         \\
                 & Rescored   1-best by BERT-text & can you let \red{doctor} know i am never using them again                                                                  \\
                 & Rescored   1-best by MATE      & can you let \blue{delta} know i am never using them again                   \\    \cline{2-3}                        
                 & Ground Truth                   & i want to {play fifa} seventeen                                                         \\
                 & Rescored   1-best by BERT-text &
                 i want to \red{leave for} seventeen                                                        \\
                 & Rescored   1-best by MATE      & i want to \blue{play fifa} seventeen 
                 \\ \cline{2-3}                        
                 & Ground Truth                   & what do you know about {fringe} in edinburgh next year                                                    \\
                 & Rescored   1-best by BERT-text &
                 what do you know about \red{french} in edinburgh next year                                                        \\
                 & Rescored   1-best by MATE      & what do you know about \blue{fringe} in edinburgh next year
                 \\ \hline
Voxpopuli        & Ground Truth                   & for example the report talks about the rule of law and corruption                          \\
                 & Rescored   1-best by BERT-text & for example the report talks about the rule of law \red{on} corruption                           \\
                 & Rescored   1-best by MATE      & for example the report talks about the rule of law \blue{and} corruption                          \\ \cline{2-3} 
                 & Ground Truth                   & i have met them they are young capable and visionary \\
                 & Rescored   1-best by BERT-text & i have met them they are young capable and \red{missionary}                                      \\
                 & Rescored   1-best by MATE      & i have met them they are young capable and \blue{visionary}\\ \hline
MTDialogue       & Ground Truth                   & it's muffled                                                                               \\
                 & Rescored   1-best by BERT-text & it's \red{muff}                                                                                  \\
                 & Rescored   1-best by MATE      & it's \blue{muffled}                                                                               \\ \cline{2-3} 
                 & Ground Truth                   & how much she got to pay                                                                    \\
                 & Rescored   1-best by BERT-text & how much \red{he} got to pay                                                                     \\
                 & Rescored   1-best by MATE      & how much \blue{she} got to pay                                                                    \\ \hline
ConvAI           & Ground Truth                   & why did the noodle box in greensborough fail its health inspection                         \\
                 & Rescored   1-best by BERT-text & why did the noodle box in \red{greensboro} fail its health inspection                            \\
                 & Rescored   1-best by MATE      & why did the noodle box in \blue{greensborough} fail its health inspection                         \\ \cline{2-3} 
                 & Ground Truth                   & tell me about duty free shopping                                                           \\
                 & Rescored   1-best by BERT-text & tell me about duty free \red{shop}                                                               \\
                 & Rescored   1-best by MATE      & tell me about duty free \blue{shopping}                                                           \\ \hline
\end{tabular}
}

\caption{Qualitative examples: We contrast the 1-best outputs of BERT-text model and \mymethod in reference to ground truth. We can observe that \mymethod improves recognition of content words and slot entities.
}
\label{tab:case}
\end{table*}

\end{document}